\documentclass[futureinternet,article,accept,moreauthors,pdftex,10pt,a4paper]{mdpi}

\usepackage{rotating}
\usepackage{chngcntr}
\usepackage{epstopdf}

\usepackage{booktabs} 
\usepackage{multirow}
\usepackage{soul}
\usepackage{microtype}

\firstpage{1} 
\articlenumber{14}
\doinum{10.3390/fi8020014}
\pubvolume{8(2)}
\pubyear{2016}
\copyrightyear{2016}
\externaleditor{Academic Editor: Emilio Ferrara}
\history{Received: 4 December 2015; Accepted: 6 April 2016; Published: 20 April 2016}

\Title{The Evolution of Wikipedia's Norm Network}

\Author{Bradi Heaberlin $^{1,2}$ and Simon DeDeo $^{1,3,4,5}$*}
\AuthorNames{Bradi Heaberlin and Simon DeDeo}

\address{%
$^{1}$ \quad Program in Cognitive Science, Indiana University, 1900 E 10th St, Bloomington, IN 47406, USA; \linebreak{bmheaber@umail.iu.edu}\\ 
$^{2}$ \quad Department of Political Science, Indiana University, 1100 E 7th St, Bloomington, IN 47405, USA\\
$^{3}$ \quad Center for Complex Networks and Systems Research, Department of Informatics, Indiana University, \mbox{919 E 10th St}, Bloomington, IN 47408, USA\\
$^{4}$ \quad Ostrom Workshop in Political Theory and Policy Analysis, 513 N Park Avenue, Bloomington, IN 47408, USA\\
$^{5}$ \quad Santa Fe Institute, 1399 Hyde Park Road, Santa Fe, NM 87501, USA\\
}

\corres{Correspondence: sdedeo@indiana.edu; Tel.: +1-812-856-2855; Fax: +1-812-856-3825}

\abstract{Social norms have traditionally been difficult to quantify. In any particular society, their sheer number and complex interdependencies often limit a system-level analysis. One exception is that of the network of norms that sustain the online Wikipedia community. We study the fifteen-year evolution of this network using the interconnected set of pages that establish, describe, and interpret the community's norms. Despite Wikipedia's reputation for \textit{ad hoc} governance, we find that its normative evolution is highly conservative. The earliest users create norms that both dominate the network and persist over time. These core norms govern both content and interpersonal interactions using abstract principles such as neutrality, verifiability, and assume good faith. As the network grows, norm neighborhoods decouple topologically from each other, while increasing in semantic coherence. Taken together, these results suggest that the evolution of Wikipedia's norm network is akin to bureaucratic systems that predate the information age.}



\begin{document}

\section{Introduction}

\noindent
A society's shared ideas about how one ``ought'' to behave govern essential features of economic and political life~\cite{sherif1936psychology,durkheim1938rules,akerlof1976economics,geertz1994thick,ellickson2009order,bowles2009microeconomics}. Outside of idealized game-theoretic environments, for example, economic incentives are supplemented with norms about honesty and a higher wage is possible when workers believe they ought not to cheat their employer~\cite{simon1951formal}. And, while the rational structure of rules and laws is an important part of coordinating actions and desires~\cite{brennan2008reason}, people determine the legitimacy of these solutions based on beliefs about fairness and authority. A police force without legitimacy cannot enforce the law~\cite{tyler2006psychological,tyler2008legitimacy}.

Norms are also under continuous development. The modern norm against physical violence, for example, has unexpected roots and continues to evolve~\cite{elias2000civilizing,pinker2011better,klingenstein2014civilizing}. Yet, we understand far less about the history and development of norms than we do about economics or the law~\cite{ehrlich2005evolution}. We often lack the data that would allow us to track the coevolution of complex, interrelated and interpretive ideas, such as honesty, fairness, and authority, the way we can track prices and monetary flows or the creation and enforcement of statutes. 

Online systems, such as Wikipedia, provide new opportunities to study the development of norms over time. Along with information and code repositories at the center of the modern global economy, such as GNU/Linux, Wikipedia is a canonical example of a knowledge \mbox{commons~\cite{hess06,benkler2006wealth,bollier06,frischmann2014governing}}. \mbox{Knowledge commons} rely on norms, rather than markets or laws, for the majority of their governance~\cite{ostrom90, hess2011understanding}. On Wikipedia, editors collaborate to write encyclopedic articles in a community-managed open source environment~\cite{west2008getting, o2007governance}, and they rely on social norms to standardize and govern their editing decisions~\cite{beschastnikh2008wikipedian}. Wikipedia's minute-by-minute server logs cover more than fifteen years of norm creation and evolution for a population of editors that has numbered in the tens of thousands. Norms matter on Wikipedia in ways that make it impossible for participants to ignore: it is the system of norms, rather than just laws, that dictates what content is or is not included, who participates, and what they do.

Paralleling findings in the study of rule evolution in large academic institutions~\cite{march2000dynamics}, we expect Wikipedia's norms to play a role in the preservation of institutional memory, to be a source of both institutional stability and change, and to bear a complex relationship to the circumstances that led to their creation. Norm pages play key roles in coordinating behavior among the encyclopedia's editors~\cite{butler2008don}. Editors commonly cite norms on article talk pages in an attempt to coordinate~\cite{schneider2010qualitative}, build consensus, and resolve disputes~\cite{kriplean2007community,beschastnikh2008wikipedian}. 

This study focuses on a subspace of the encyclopedia devoted to information and discussion about the norms of the encyclopedia itself. The communities associated with each of Wikipedia's 291 languages and editions have a great deal of independence to define and change the norms they use; thus, each can follow a different evolutionary trajectory. Here, we focus solely on norms in the English-language Wikipedia. We study the evolution of these norms using a subset of tightly-linked pages that establish, describe, and interpret them. These pages, along with the relationships between them, allow us to quantify how editors describe expectations for behavior and, consequently, how they create and reinterpret the norms of \mbox{their community}.

We focus on the links between norm pages. Online link formation occurs for a variety of reasons~\cite{park2003hyperlink}, including strategic association by the individual making the citation~\cite{gonzalez2009opening}. In the case of Wikipedia, links between pages in the encyclopedia ``mainspace'' encode information about semantic relationships~\cite{strube2006wikirelate,witten2008effective} and the relative importance of pages~\cite{bellomi2005network,lizorkin2009analysis}. Extending these analyses to the norm pages of the encyclopedia allows us to see how norms are described, justified, and explained by reference to other norms. Our use of this network parallels studies of citations in legal systems; researchers use legal citations to track influence via precedence~\cite{fowler2008authority} and legitimation~\cite{walsh1997meaning}, as well as the prestige of the cited~\cite{caldeira1985transmission, walsh1997meaning}. The parallel to legal citations is not exact: the pages within Wikipedia's norm network are not (usually) created in response to a particular event, as in a court case, but rather in response to a perceived need; pages can be created by any user, rather than a particular judge or court; and pages can be retrospectively edited (leading, for example, to the potential for graph cycles when new links \mbox{are introduced}).

This network perspective allows us to go beyond the tracking of a single behavior over time (\mbox{a common} approach in studies of cultural evolution~\cite{henrich2008five}) to look at the evolution of relationships between hundreds, and even thousands, of distinct ideas. We use these data to ask three critical questions. In a system where norms are constantly being discussed and created, how and when do some norms come to dominate over others? What types of behavior do they govern? Additionally, how do those core norms evolve over time?

The answers are surprising. While some accounts of Wikipedia stress its flexibility and the \textit{ad hoc} nature of its governance~\cite{shirky2008here,SOCF:SOCF1090,konieczny2010adhocratic}, we find that Wikipedia's normative evolution is highly conservative. Norms that dominate the system in Wikipedia's later years were created early, when the population was much smaller. These core norms tell editors how to write and format articles; they also describe how to collaborate with others when faced with disagreements and even heated arguments. \mbox{To do} this, the core norms reference universal, rationalized principles, such as neutrality, verifiability, civility, and consensus. Over time, the network neighborhoods of these norms decouple topologically. \mbox{As they} do so, their internal semantic coherence rises, as measured using a topic model of the page text. Wikipedia's abstract core norms and decoupling process show that it adopts an ``institutionalized organization'' structure akin to bureaucratic systems that predate the information age~\cite{meyer77}.

\section{Methods}

To gather data on the network of norms on Wikipedia, we spider links within the ``namespace'' reserved for (among other things) policies, guidelines, processes, and discussion. These pages can be identified because they carry the special prefix ``Wikipedia:'' or ``WP:''. Network nodes are pages. Directed edges between pages occur when one page links to another via at least one hyperlink that meets our filtering criteria; these links are found by parsing the raw HTML of each page and excluding standard navigational templates and lists. Our network is thus both directed and unweighted. We begin our spidering at the (arbitrarily selected) norm page ``Assume good faith''. Details of the spidering process, hyperlink filters and our post-processing of links between pages appear in Appendix~\ref{appendix_a}; both the raw data and our processed network are freely available online~\cite{online_ref}.

Editors classify pages in the namespace by adding tags; these tags include, most notably, ``policy'', ``guideline'', and ``essay'', among others. When we download page text, we also record these categorizations. These categorizations describe gradated levels of expectations for adherence~\cite{morgan2010negotiating}. In automatically-included ``template'' text, policies are described as ``widely accepted standards'' that ``all editors should normally follow''~\cite{policy_template}, guidelines as ``generally accepted standards'' that ``editors should attempt to follow'' and for which ``occasional exceptions may apply''~\cite{guideline_template}, while essays provide ``advice or opinions'': ``[s]ome 
essays represent widespread norms,'' while ``others only represent minority viewpoints''~\cite{essay_template}. A fourth category is the ``proposal'', which describes potential policies and guidelines ``still ... in development, under discussion, or in the process of gathering consensus for adoption''~\cite{proposal_template}.

Previous analysis of Wikipedia's policy environment has emphasized the many, often overlapping, functions that norms play in the encyclopedia, such as policies that both attempt to control un-permitted use of copyrighted material and to establish legitimacy through the use of legal diction and grammar~\cite{butler2008don}. In the current study, we consider a complementary classification system that focuses on the types of interactions the norms govern, rather than their functions. We propose three distinct \mbox{norm categories} based on, and extending, pre-existing classification of the norms that govern natural~\cite{ostrom90} and knowledge commons~\cite{hess2011understanding}. 

Norms may attempt to regulate content creation (``user-content'' norms) and interactions between users (``user-user'' norms). In addition, norms may attempt to define a more formal administrative structure with distinct roles, duties, and expectations for admins (``user-admin'' norms). \mbox{The two authors} of this paper independently categorized a random sample of forty pages using this scheme, and we calculated inter-coder reliability using Cohen's kappa~\cite{cohen1960coefficient}.

For our semantic analysis, we include all text, except that found in special boxes whose text is replicated by template across multiple pages. To build our distribution over one-grams, we normalize all text to lowercase, merge hyphenated words (``error-correction'' to ``errorcorrection''), and drop punctuation (``don't'' to ``dont''). We do neither stemming nor spelling correction.

A critical external variable is the number of active users on the encyclopedia at any point in time. Following \cite{halfaker2012rise}, we define an active user as one who has made five or more edits within a month; these statistics are publicly maintained at \cite{wiki_stats}.

\subsection{Centrality and Attention Measures}

The pages in our corpus are created to explain the norms of Wikipedia to editors and influence their interactions with the encyclopedia's editing community and content. Users navigate the system of norms as a network structure and consequently encounter some pages more than others.
 
We measure this using eigenvector centrality (EC), which quantifies the importance of a page based on its overall accessibility within the network. The EC of a page is the probability of happening across a page during a random walk; equivalent to the PageRank 
 algorithm, it is used in the behavioral sciences to identify consensus on dominance and power~\cite{ec_jess}. We set $\epsilon$, the probability of a random jump, to 0.15.

We expect some pages to become highly central to the network, while others remain largely peripheral. We quantify the inequality of the system using the Gini coefficient (GC). GC varies between zero (perfect equality; all pages have equal EC) and one (one page has a high EC; all other pages have the same low value). GC is widely used in economics to measure income inequality. \mbox{Here, it} provides a global measure of the extent to which a system is dominated by a few norms. As a dimensionless quantity, it allows researchers to compare this system to others that might be the subject of later research.

Because we are interested in the ways in which the norm citation network evolves and the role that norms play in the context of this structure, EC is an ideal measure of a norm's importance. \mbox{In addition} to quantifying structural importance, however, we expect EC to correlate with, and to predict, behavioral measures of the attention a page receives. To measure the relationship between centrality and behavioral measures of attention, we track page view data (from Wikipedia's server logs made available by StatsGrok~\cite{stats_grok}, see Appendix~\ref{page_views}), the total number of edits a page has received, the number of edits on its associated talk page, and the number of editors who have edited the page. We perform a multivariate linear regression on these attention measures, along with page age and page size (\mbox{in bytes}) as predictors of a page's EC (see Appendix~\ref{regression_section}).

\subsection{Influence and Overlap}

An important feature of the norm network is the sphere of influence: the pages that rely on any particular page for context.

Consider, for example, the norm page ``Neutral Point of View'' (NPOV), a page urging editors to describe article subjects without taking sides. A page that links to NPOV relates its own subject to NPOV in some fashion. For example, among many pages that link to NPOV is ``Propaganda'', an essay urging editors to be wary of using propaganda outlets of authoritarian governments. The Propaganda page links to the NPOV page in order to define the notion of ``undue weight''; NPOV's content can thus be said to influence the interpretation of what is found on Propaganda. 

Influence is distinct from centrality; centrality measures the extent to which pages link to the page in question. Conversely, influence measures the extent to which the content of that page influences other pages. In our formalism, a node $p$ can be understood to \emph{influence} 
 a node $q$ when $q$ links to $p$. Influence need not be direct, however: $p$ can influence $q$ if $q$ links to $r$ and $r$ links to $p$. \mbox{To measure} the non-local influence, we consider random walks on the direction-reversed network.

More formally, placing a random-walker at node $p$, we allow her to take $n$ steps from this starting point along the direction-reversed network; we write the resulting probability distribution over the final position as $p_i$, the probability of the walker ending up at node $i$. The distribution $p_i$ defines the influence that $p$ has on $i$.

To quantify the distance between two nodes, we then consider the influence \textit{overlap} 
 between \mbox{two arbitrary} nodes $p$ and $q$. Overlap quantifies the extent to which two random walkers, beginning at these nodes, will tend to visit the same pages. If $p_i$ and $q_i$ are the probability distributions associated with the influence of node $p$ and $q$, then overlap is defined as:
\begin{equation}
O(p, q) = \frac{\sum_{i=1}^N p_i q_i}{\left(\left[\sum_{i=1}^N p^2_i\right] \left[\sum_{i=1}^N q^2_i\right]\right)^{1/2}}
\end{equation}

For multiple pages, we can compute the average pairwise overlap simply by averaging the overlap between all possible pairs within the set.

High overlap between $p$ and $q$ indicates that two pages influence a large number of common nodes. When $n$ goes to infinity, the random walkers converge to the stationary distribution, and the overlap is one; conversely, when $n$ is small, random walkers have less time to encounter each other. We take $n$ equal to five, larger than the average shortest path (roughly three, in our network), so that nodes are potentially reachable, but much less than the convergence time to the stationary distribution.

Overlap can be thought of as a measure of the separation of spheres of influence. It invokes only local mechanisms: users traveling from one page to another by the links that connect them. This is in contrast to a measure, such as shortest paths, which is computationally expensive and requires detailed, global knowledge of the network link-structure. In general, for example, the number of nodes an algorithm needs to visit in order to determine the shortest path between two nodes will usually be much larger than the length of the final path. 

Both influence and overlap require us to specify particular nodes of interest; we focus in this work on pairs of high-EC pages, or core norms.

\subsection{Semantic Coherence}

We consider the semantic relationships between pages. This provides a notion of relatedness that is distinct from how norms connect via hyperlinks. To do this, we do topic-modeling (latent Dirichlet allocation~\cite{Blei2003}) on the one-grams of the visible, human-readable text on each page. Topic models allow us to represent short texts even when they draw from a rich vocabulary: topics coarse-grain the underlying distributions over words. 

With the resulting topic model, we can then compute the semantic distance between all pairs of pages using the Jensen--Shannon distance (JSD), a measure that quantifies the distinguishability of \mbox{two distributions}~\cite{dedeo2013bootstrap}. This gives us a weighted semantic network that we can compare to the network of hyperlinks between pages. In particular, we can compute the \emph{semantic coherence}
: the Pearson correlation between $p_i$ (the influence of node $p$ on node $i$) and the negative JSD from node $p$ to node $i$, $J_i$. When nodes that are closely related topologically are also closely related semantically (JSD low), the coherence is high.

\subsection{Community Detection}

We expect the links that editors make at the local level to give rise to distinct clusters, or \textit{norm bundles}, at the global level. We use the Louvain community detection algorithm~\cite{blondel08} to detect clustering among the nodes in the network. The Louvain algorithm maximizes the modularity at each local partition of the network. The algorithm first assigns each node $i$ to a different cluster, then computes the potential modularity gain to $i$ for joining the cluster of its neighbor node $j$. Each $i$ will join the cluster of $j$ when the merge offers the highest positive modularity gain. If there is no possible gain in modularity, $i$ remains in its initial cluster.

\section{Results}

At first, Wikipedia's population underwent exponential growth. In mid-2007,~however, population growth stalled and entered a period of secular decline~\cite{halfaker2012rise}; see Figure~\ref{pop}. Over the course of this rapid growth and longer timescale decay, users created a large number of pages establishing, describing, and interpreting community norms. Our analysis finds a total of 1976 pages associated with norms. \mbox{There are} 17,235 edges between these nodes; the network density, $0.0044$, is of the same order of magnitude as those seen for academic citation networks~\cite{yan2012scholarly}; 1872 (95\%) of these pages are linked together in a giant component. 

There are a total of 56 pages classified as policy and 113 marked as guideline; for concision, we refer to pages of both types as ``policy''. \mbox{The majority} of non-policy pages (1807) are classified as ``essays'' (1255), followed by ``proposals'' (182) (suggestions either rejected by the community or under discussion), and ``humor'' pages similar to essays, but taking a more irreverent tone (125).

We were able to achieve good, but not perfect, agreement in categorizing pages as~user-content, \textls[-10]{user-user, or user-admin norms.~Our categorization agreement rate was 75\% over forty randomly-selected} pages. This is well above chance ($p\ll10^{-3}$), with Cohen's $\kappa$ value, of $0.59$ indicating ``moderate'' agreement~\cite{landis1977measurement}. We disagreed, for example, on ``Editors\_should\_be\_logged-in\_users\_(failed\_proposal)'' (user-user \emph{vs.} user-content) and ``Paid\_editor's\_bill\_of\_rights'' (user-user \emph{vs.} user-admin). In the same sample of forty random pages, we encountered only one that we believed was not a norm, giving an approximate precision rate of 97.5\%.

\subsection{Network Construction}

\begin{figure}[H]
\centering
\includegraphics[width=4.5in]{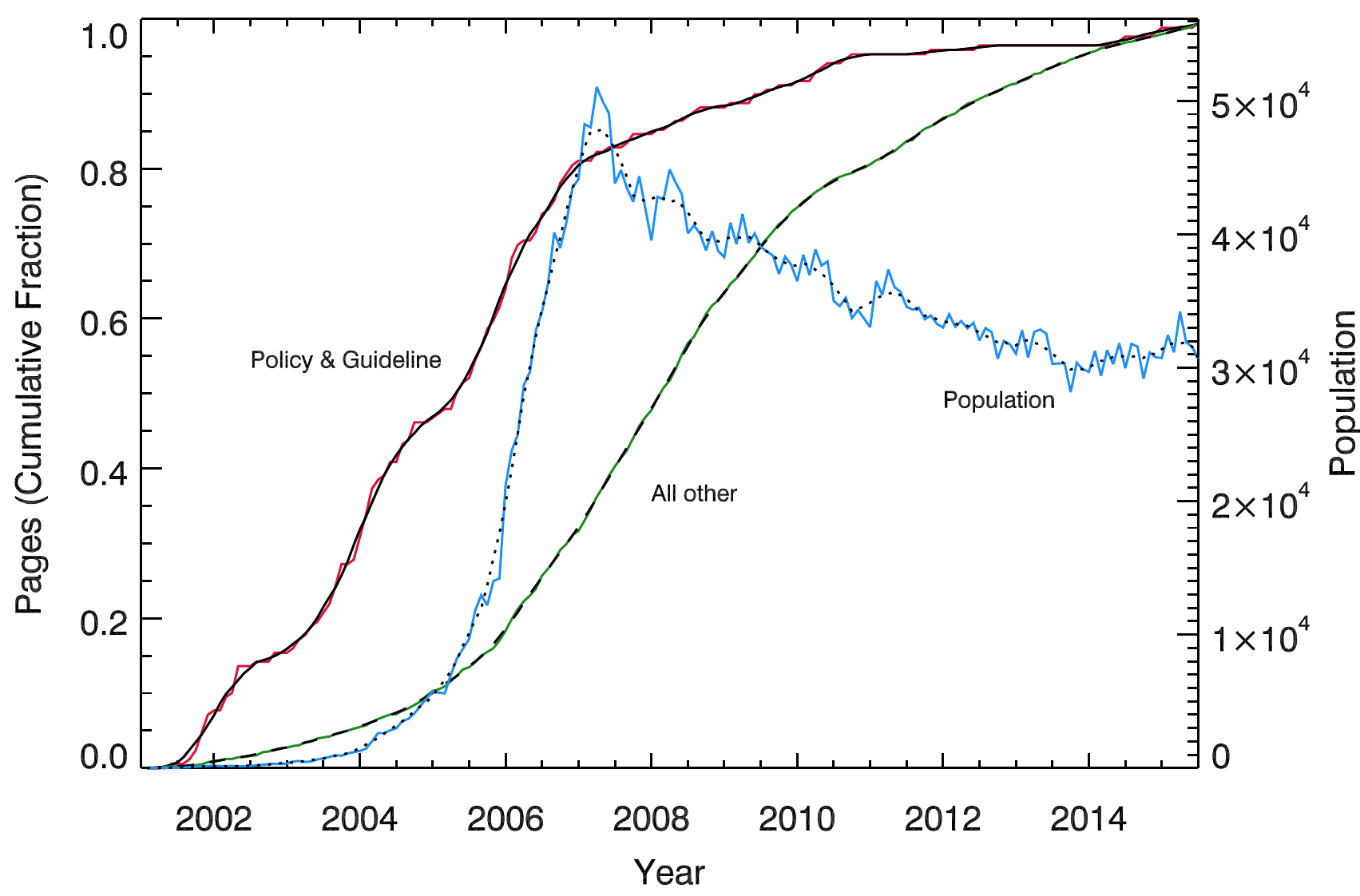}
\caption{Cumulative growth in policy (red/solid line) and non-policy (green/dashed line) pages, overlaid on active population (blue/dotted line). Policy creation precedes the arrival of the majority of users, while the creation of non-policy pages, usually in the form of essay and commentary, lags the growth in population. \label{pop}}
\end{figure} 
Most policy pages appear before the bulk of the population arrives: over half the policy pages were created by May 2005, before the population reached 20\% of its maximum. By the time the population did reach its maximum, in March of 2007, over 80\% of the policy pages had already been created. By contrast, the creation of non-policy pages in the form of essays and commentary lags population growth. When the population reached its March 2007 maximum, less than one-third of the non-policy pages were in place. It was not until a year later that half of the policy pages were in place. \mbox{This is} shown in Figure~\ref{pop}.

Eigenvector centrality leads to a distinct hierarchy of pages, with some gaining a significant fraction of the overall centrality in the system. This is shown in Appendix~\ref{scree_app}, Figure~\ref{ec}, broken out by four main page categories---policies, guidelines, essays, and proposals. Policies and guidelines dominate the system by centrality. Our centrality measure correlates with all of the of behavioral measures of attention we consider (see Appendix~\ref{page_views}, Table~\ref{correlate}). 

The hierarchy is established early and yet is remarkably stable over the lifetime of the system. \mbox{The Pearson} correlation between the eigenvector centrality of nodes in 2001 and their final values in 2015 is 0.87; year to year, it is always greater than 0.9. The growth in nodes' in-degree is roughly multiplicative; for nodes with degree less than one-hundred (93\% of the total network), the growth rate is, on average, $12.7\%\pm0.3\%$ from one year to the next. There is some evidence for super-multiplicative returns to scale; the yearly growth rate for pages with in-degree less than ten is only $10.6\%\pm0.4\%$. 

\begin{figure}[H]
\centering
\includegraphics[width=4.5in]{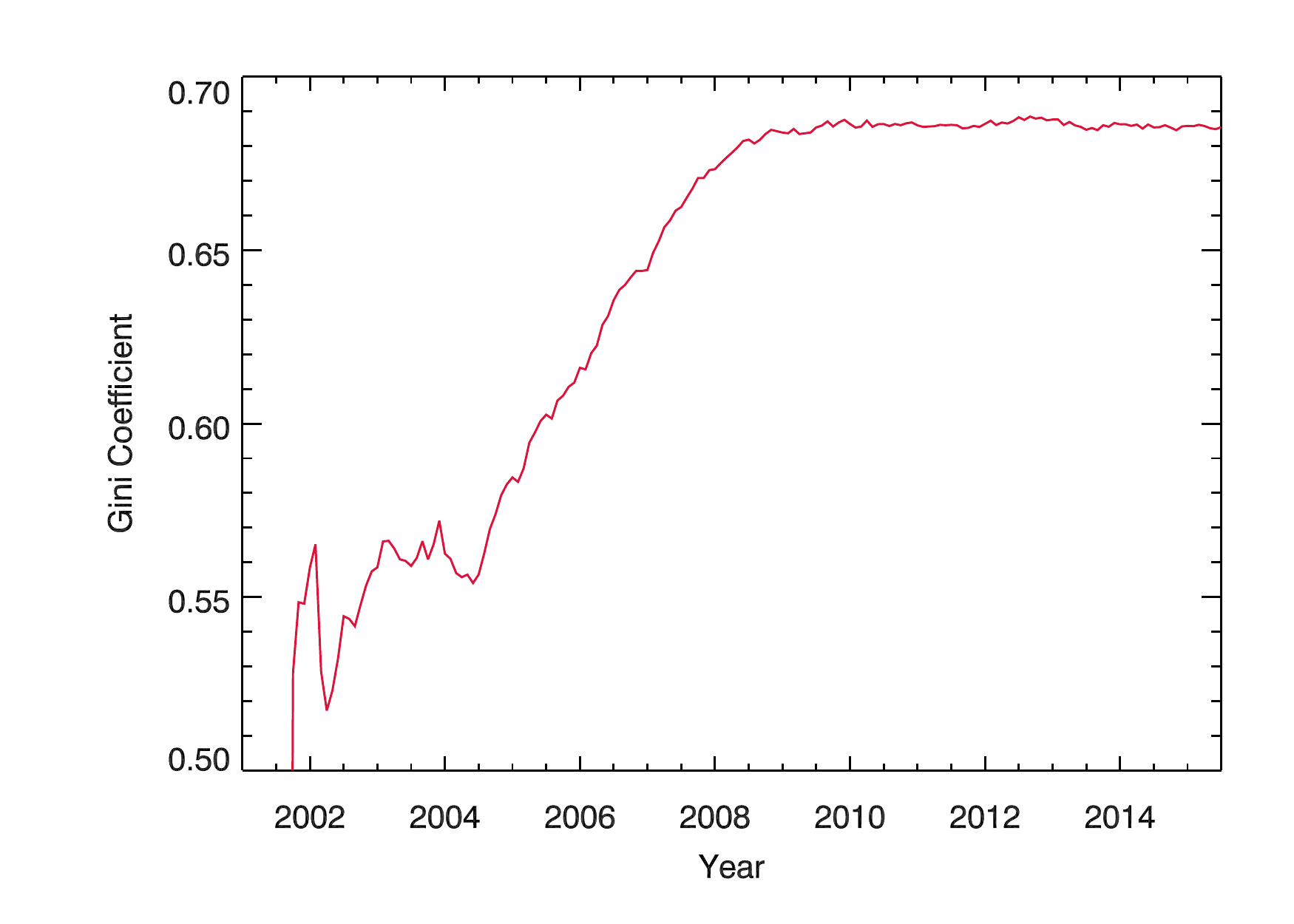}
\caption{Evolution of the Gini coefficient over time. As new pages enter the system, overall network inequality increases, stabilizing in 2008.\label{gini}}
\end{figure}

All of this means that, as new pages enter the system, they fail to gain the prominence of the early core norms. This leads to an increase in overall network inequality, shown in Figure~\ref{gini}.

In short, policy growth precedes population growth. Policies have far greater centrality in the network than other page types. Centrality in the network is unequally distributed and becomes less equal over time. 

\subsection{Core Norms}

Table~\ref{ec_table} lists the top twenty pages in our network. These core norms govern a range of behaviors, including user-content actions (write articles from a neutral point of view, \#1; include only verifiable information, \#2; and reliable sources, \#3), user-user actions (find consensus, \#6; assume good faith, \#11; be civil, \#16; do not ``edit war'', \#19), and user-admin relationships involving specially-defined roles (blocking policy, \#13; the arbitration committee, \#17). In some cases, a norm spans multiple classes; ``What Wikipedia is not'', for example, includes both ``Wikipedia is not a dictionary'' (a norm on the nature of the content to be included) and ``Wikipedia is not a battleground'' (a norm on how users should interact with each other).

\begin{table}[H]
\centering
\begin{tabular}{clll} 
\toprule
\textbf{Rank} & \textbf{Name} & \textbf{Classification} & \textbf{Creation Date} \\ \midrule
1 & Neutral\_point\_of\_view & User-content & 24 December 2001 \\ 
2 & Verifiability & User-content & 2 August 2003 \\ 
3 & Identifying\_reliable\_sources & User-content & 28 February 2005 \\ 
4 & What\_Wikipedia\_is\_not & User-user/user-content & 24 September 2001 \\ 
5 & Biographies\_of\_living\_persons & User-content & 17 December 2005 \\ 
6 & Consensus & User-user & 11 July 2004 \\ 
7 & Policies\_and\_guidelines & User-user/user-content & 1 November 2001 \\ 
8 & Administrators & User-admin & 16 May 2001 \\ 
9 & No\_original\_research & User-content & 21 December 2003 \\ 
10 & Citing\_sources & User-content &19 April 2002 \\ 
11 & Assume\_good\_faith & User-user & 3 March 2004 \\ 
12 & Notability & User-content & 7 September 2006 \\ 
13 & Blocking\_policy & User-admin & 8 June 2003 \\ 
14 & Dispute\_resolution & User-user/user-admin & 12 January 2004 \\ 
15 & Redirect & User-content & 25 February 2002 \\ 
16 & Civility & User-user & 5 February 2004 \\ 
17 & Arbitration\_Committee & User-admin & 16 January 2004 \\ 
18 & Vandalism & User-content & 29 March 2002 \\ 
19 & Edit\_warring & User-user & 26 April 2003 \\ 
20 & Talk\_page\_guidelines & User-user & 15 April 2005 \\ 
\bottomrule
\end{tabular}
\caption{Core norms. Top twenty pages, by eigenvector centrality, in 2015. All are either policy or guideline pages, and all were in place by the end of 2006. The majority of these core norms were created before 2004, when the population was less than 3\% of its peak.\label{ec_table}}
\end{table}

All of these core norms were created early in the system's history. The majority were created before 2004, when the population was less than 3\% of the March 2007 peak. The earliest members of the community first defined and articulated its core norms.

It is important to note that while the most important norms are those that are created early, not all of the pages created early become, or remain, central to the network. This is shown visually in Appendix~\ref{regression_section}, Figure~\ref{regression_explain}; there are many old pages that never grew to importance and that have ECs comparable to the youngest pages. Because of this, page age alone is not a significant predictor of eigenvector centrality. We confirm this with a multivariate linear regression (see Table~\ref{regression_table}). \mbox{The number of} editors is a strong predictor; not only do high EC pages attract a large number of unique editors, but there are few low-EC pages that do.

\subsection{Overlap and Semantic Coherence}

Over the course of network construction, core norms are drawn apart topologically. At the same time, the semantic coherence of their neighborhoods rises.

\begin{figure}[H]
\centering
\includegraphics[width=4.5in]{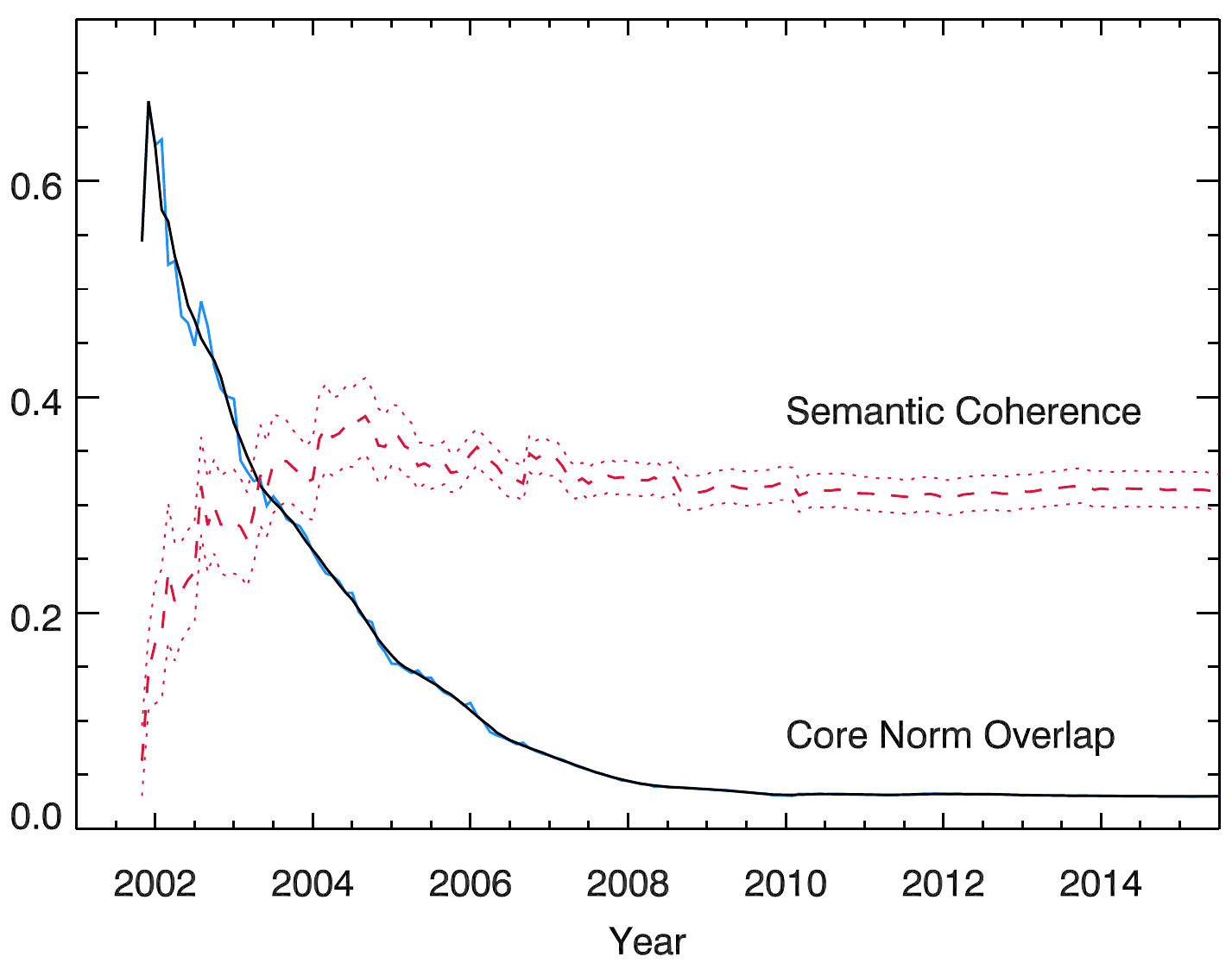}
\caption{Evolution of influence overlap among the core norms (top twenty norms by eigenvector centrality) over time (solid line, labeled). In terms of the pages they influence, core norms draw apart over time, stabilizing in 2008. At the same time, semantic coherence (dashed line, labeled) increases. Neighborhoods become topologically distinct, but internally coherent.\label{overlap}}
\end{figure}

Figure~\ref{overlap} shows the average pairwise overlap between the top ten pages in our network (since some norms are created later, the number of norms in this final set is lower early on). Early in the system history, when the network is small, overlap is very high. The creation of new pages leads to \mbox{a rapid} decline in overlap; even in 2006, when all core norms are in place, the overlap continues to decline. Figure~\ref{overlap} also shows the evolution of semantic coherence, which rises rapidly and \mbox{stabilizes early}.

Network growth could have been imagined to drive a knitting together of distinct principles. Instead, the opposite happens: core norms slowly draw apart as page creation leads to distinct spheres of influence. Rather than nucleating around a set of densely-connected core principles, the norm network continues to condense around multiple points.

We note that the local clustering coefficient, a measure of the extent to which two nodes, linked to the same node, tend to also link together, remains essentially constant over the span of the data (\mbox{see Appendix}~\ref{app_clustering}, Figure~\ref{clustering}). The ways in which editors link together small groups of pages changes little while their cumulative effect produces large and lasting changes both in attention inequality and page overlap.

\subsection{Emergent Clusters}

\begin{figure}[H]
\begin{tabular}{cc}
\includegraphics[width=2.9in]{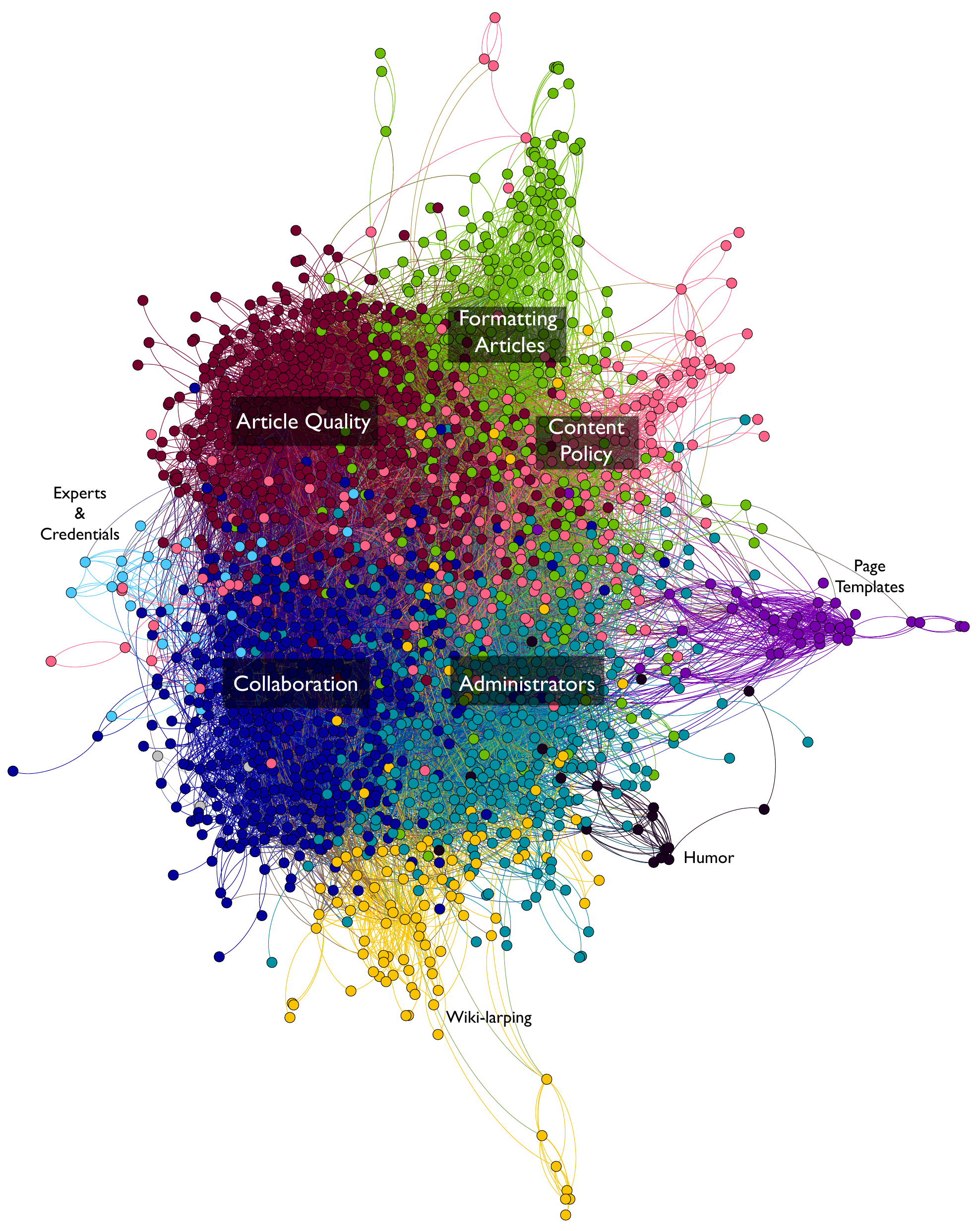} & \includegraphics[width=2.9in]{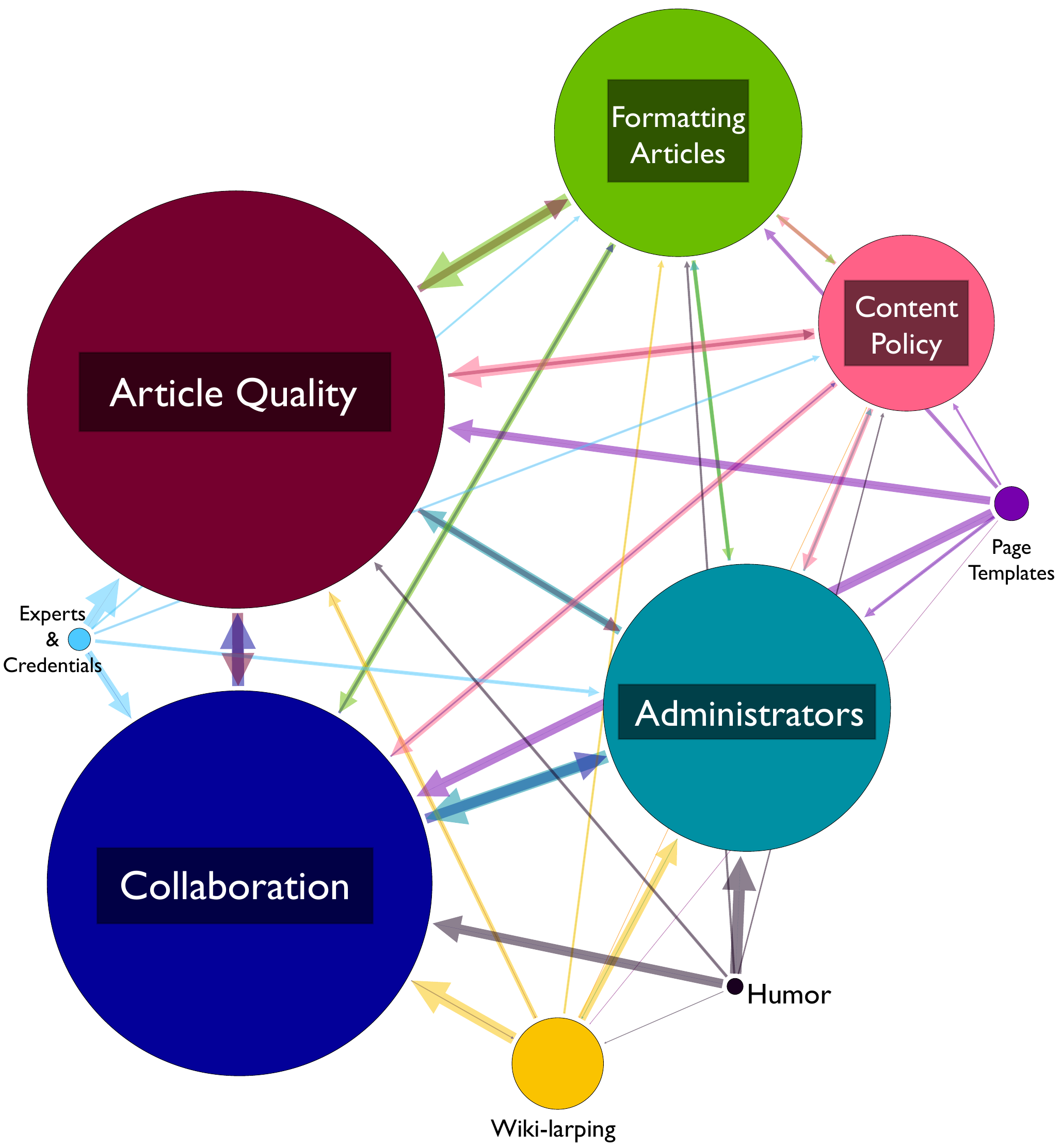}
\end{tabular}
\caption{The topology of the norm network is organized around five central clusters, found using the Louvain algorithm. Cluster themes are based on a sample of high-eigenvector centrality (EC) nodes in each cluster and confirmed by reference to a topic model of word usage. Left panel: full network, with cluster membership indicated by color. Right panel: cluster structure. Each node is a Louvain cluster, and node size indicates cluster size by number of pages. Edge weights are defined as the fraction of the origin cluster's out-links that link to each other cluster (self-loops are not shown). \label{louvain}} 
\end{figure}

The connected component of network, containing 95\% of all nodes, partitions into 10 clusters. In Table~\ref{community_table}, we describe the top nine, which together nearly all of the giant component. By inspecting the top ten nodes in each cluster, we classify them into user-content, user-user, and user-admin norms (see Table~\ref{topic_ec}). A force-directed layout (ForceAtlas2, implemented in Gephi~\cite{forceatlas}) allows us to visualize the norm network and the topological relationships between its emergent groups (see Figure~\ref{louvain}). 

\begin{table}[H]
\centering
\begin{tabular}{ccll}
\toprule
\textbf{Rank} & \textbf{Fraction of System} & \textbf{Classification} & \textbf{Topic} \\ \midrule 
1 & 24.8\% & User-Content & Article Quality\\ 
2 & 22.9\% & User-User & Collaboration\\
3 & 17.1\% & User-Administration & Administrators\\
4 & 14.7\% & User-Content & Formatting Articles\\
5 & 10.5\% & User-Content & Content Policies\\
6 & 5.4\% & User-User & Wiki-larping
\\
7 & 2.0\% & User-Content & Page Templates\\
8 & 1.3\% & User-User/User-Content & Experts and Credentials\\
9 & 1.0\% & User-User & Humor\\
\bottomrule
\end{tabular}
\caption{Top nine Louvain clusters, by number of nodes. Communities fall into three classifications (user-user, user-content, user-administration), based on the interactions they govern; we determine these labels by inspecting the top ten nodes by centrality within each cluster.\label{community_table}}
\end{table}

The five largest clusters comprise roughly 90\% of the network. The Article Quality cluster includes nodes such as Neutral Point of View, Verifiability, and Reliable Sources, governing how articles should be written. The Collaboration cluster includes pages on Consensus, Assume Good Faith, and Edit Warring, describing policies and norms associated with interpersonal interaction. \mbox{The Administrators} cluster contains pages relevant to administrative actions, such as the Blocking Policy and the Arbitration Committee. The Formatting cluster contains articles such as Manual of Style, Article Titles, and Disambiguation. Additionally, the Content Policies cluster contains articles on copyrights, copyright violations, and policies on image use and use of non-free content. The remaining clusters include a small group of articles on page templates; one on the role of experts of Wikipedia; two groups of humor pages (Wiki-larping, a humorous take on Wikipedia as if it were a Dungeons and Dragons game, and a cluster of pages, including ``Bad Jokes and Other Deleted Nonsense'').

Each of the top nine clusters is associated with a distinct topic in our topic model (see Appendix~\ref{k20}, Table~\ref{topic_list}); while the article quality cluster is the largest by node number, the topic associated with the collaboration cluster dominates the system by word. Even task-based norms appear to draw on the semantics of interpersonal cooperation. 

\section{Discussion}

The most influential pages in the norm network are also the earliest to be created. A Matthew effect~\cite{merton1968matthew} appears to operate for social norms, where later additions to the network do not grow in influence quickly enough to destabilize the hierarchy. Why are there no normative revolutions \mbox{on Wikipedia}?

Perhaps the earliest users know best: their policies work well and are simply adopted by those who come later; or, later users may join precisely because they subscribe to the norms that have already been articulated. Users who disagree with these norms may find that reinterpretation, rather than replacement, is a more effective response given the disproportionate allocation of attention to early pages.

The fact that core norms are created so early means that a relatively small number of users set them in place. This group may have created norms that meet their own needs, but not the needs of those who arrive later. For example, if early users are predominantly university students with flexible working hours, for example, they may develop norms that implicitly rely on the possibility of responding to criticism in short, rapid bursts. If later arrivals do not have the same flexibility, but the norms persist, they will find themselves at a relative disadvantage in conflicts that arise, even if the amount of effort they devote to the system each week is the same.

Recent work~\cite{shaw14} has suggested that early users later form an oligarchy that monopolizes power, subverts democratic control, and comes into increasing conflict with the larger collective. If this is true, the enduring centrality of their own interests in the norm network may be \mbox{a source} of power.

Alternatively, the influence of a small group of editors may persist via the core norms despite a gradual decentralization of power within the encyclopedia. One ethnographic account of Wikipedia's editing community~\cite{forte2009decentralization} suggests that a group of ``old-timers'' brings important social norms with them into the encyclopedia's increasingly local governance structures, such as WikiProject communities. Our findings show that the structure of the norm network is, by measures of page count, clustering, core norm overlap, and semantic coherence, largely stable by 2008. Thus, the core norms and global norm structure analyzed here may provide an early foundation of norms for small, decentralized communities that form later in the \mbox{encyclopedia's development}.

Much of Wikipedia's network simply coordinates technical practices, such as article naming conventions. The most important norms, however, attempt to rationalize the system around universal concepts, such as neutrality, verifiability, consensus, and civility. An important insight comes from a theory of bureaucracy and institutionalized organization developed by Meyer and Rowan (1977~\cite{meyer77}). They propose that norms such as these can function as institutional myths that make the system appear legitimate and less \emph{ad hoc}, by connecting it to a rational framework. 

Page creation continues to grow long after the core norms are already in place. What happens when editors continue to develop and refine this network?

Meyer and Rowan's theory predicts the phenomenon of decoupling, driven by the emergence of inconsistencies between different myths. The essay Civil\_POV\_pushing, for example, describes how some users may be able to violate the neutrality norm by strict adherence to norms of civility. In Meyer and Rowan's theory, pages like these, that attempt to resolve inconsistencies between myths, will be rare. Myths will instead tend to decouple from each other over time. 

Our quantitative findings are consistent with this prediction.~As the system grows, the creation of norm-spanning pages, such as Civil\_POV\_pushing, are rare and insufficient to prevent the neighborhoods of the core norms drawing apart into separate spheres of influence with high internal semantic coherence. In successful systems, decoupling is also expected to happen not only between myths, but between these myths and actual practice, a phenomenon pointed to by the existence of the page ``Ignore\_all\_rules'' (``if a rule prevents you from improving Wikipedia, ignore it'').

Our findings are also consistent with Meyer and Rowan's second major prediction: that systems become increasingly reliant on a logic of good faith rather than following procedure. Not only is ``Assume good faith'' itself a core norm, but the associated topic dominates the system as a whole.

The norm network we study here is the culmination of over thirty thousand edits. We analyze the development of this system over time via the editing community's collective decisions and their allocation of attention within the network. While this method tells us a great deal about the collective process of norm creation, we do not know how individual editors understand the relationships between norms or use them to guide how they edit and interact with others. Rather than memorize the complex network in its entirety, an editor may coarse-grain its properties to form his or her own mental representation of the encyclopedia's normative structure. Editors' mental representations might then inform their linking and editing behaviors, creating a feedback loop between the representation and the norm network as a whole.

\section{Conclusions}

Norms are a crucial unit of cultural evolution, and they gain meaning and force from the relationships that connect them. Our work here has studied the evolution, over fifteen years, of the interdependent network of norms at the center of Wikipedia.

The evolution of this network is a remarkably conservative process. Early features are maintained, and in some cases even amplified, over the course of the network's development. Our findings are consistent with the ``iron law'' of oligarchy in peer-production systems; they also complement accounts of gradual decentralization in Wikipedia's governance structure.

The encyclopedia's core norms address universal principles, such as neutrality, verifiability, civility, and consensus. The ambiguity and interpretability of these abstract concepts may drive them to decouple from each other over time. Wikipedia is a paradigmatic example of a 21st Century knowledge commons. Yet, its core norms play a structural role analogous to the institutional myths of rationalized 20th Century bureaucracies.

\vspace{6pt}
\acknowledgments{We thank John Miller (Carnegie Mellon), Stephen Benard (Indiana University) and \mbox{Cris Moore} (Santa Fe Institute) for helpful discussions, and the Santa Fe Institute for their hospitality when this work was begun. Bradi Heaberlin was supported by the Research Experience for Undergraduates program 
 at the Santa Fe Institute under National Science Foundation Award \#ACI-1358567, by the Cox Research Scholarship Program and by the Indiana University Science, Technology and Research Scholars (STARS) program. Data used in this analysis are available online~\cite{online_ref}.} 

\authorcontributions{The authors jointly designed the research concept, gathered the data, and conducted the analyses. Both authors jointly wrote the paper.}

\conflictofinterests{The authors declare no conflict of interest.} 

\clearpage



\appendix
\counterwithin{figure}{section}
\counterwithin{table}{section}

\section{Corpus Construction}
\label{appendix_a}

As described in the main text, we build our corpus by spidering outward from the page ``Assume good faith'', following all links in the Wikipedia namespace to build a directed, unweighted network. Not all pages within the namespace are normative, however. After completing the spidering process, we remove pages that are solely lists (e.g., the pages ``List of guidelines'' or ``Lists of protected pages'') that describe ``projects'', or other initiatives focused on outreach (e.g., ``Wikipedia Loves Libraries''), or on adding a certain kind of content to the encyclopedia (e.g., ``WikiProject Libertarianism''), or that serve as noticeboards (e.g., the ``Village pump'', ``Media copyright questions''), with filters on both page titles and editor-assigned categories.

Many page names have synonyms (e.g., ``AGF'' redirects to ``Assume good faith''); we merge synonyms. Not all links between pages indicate a deliberate decision to connect one norm to another. Many pages, for example, contain ``boxes'', small code snippets that categorize pages or provide navigation indices to similar norms. These boxes can be created by a single command and are replicated across multiple pages; we do not include out-bound links found in these boxes. We do not count multiple links between pages; our edges are unweighted; a directed edge between A and B refers only to the presence of at least one link from A to B. Pages sometimes have internal links; we drop all self-edges. Our spidering includes only pages that existed on 12:00:00 UTC, 20 August 2015. 

\section{Relationship between Eigenvector Centrality and Attention Measures}
\label{page_views}

To compare the norm network structure and user attention, we measure the correlation between the centrality of a norm page and the percent of the network's page views that the norm accumulates over a 31-day period (July 2015). We find a moderate correlation~\cite{cohen1992power}, $r = 0.32$, between EC and page views. (The distribution of EC and page view values is slightly non-linear. We conduct a power-law fit and find that $\alpha = 1.42 \pm 0.02$. Consequently, a page that doubles its EC more than doubles its share of the network's page views. For simplicity in this analysis, we present the linear correlations.)

EC correlates significantly with all behavioral attention measures we consider; not just page views, but number of edits, number of talk page edits and number of editors; see Table~\ref{correlate}. 
\begin{table}[H]
\centering
\begin{tabular}{lcc}
\toprule
\textbf{Attention Measure} & $\textbf{\emph{r}}$ & $\textbf{\emph{p}}$ \textbf{Value} \\ \hline
Page views & 0.32 & $<$$10^{-3}$ \\
Number of edits & 0.70 & $<$$10^{-3}$\\ 
Number of talk page edits & 0.63 & $<$$10^{-3}$ \\
Number of editors & 0.72 & $<$$10^{-3}$ \\
\bottomrule
\end{tabular}
\caption{Correlation of eigenvector centrality with behavioral measures of attention.\label{correlate}} 
\end{table}


\begin{figure}[H]
\centering
\includegraphics[width=3.5in]{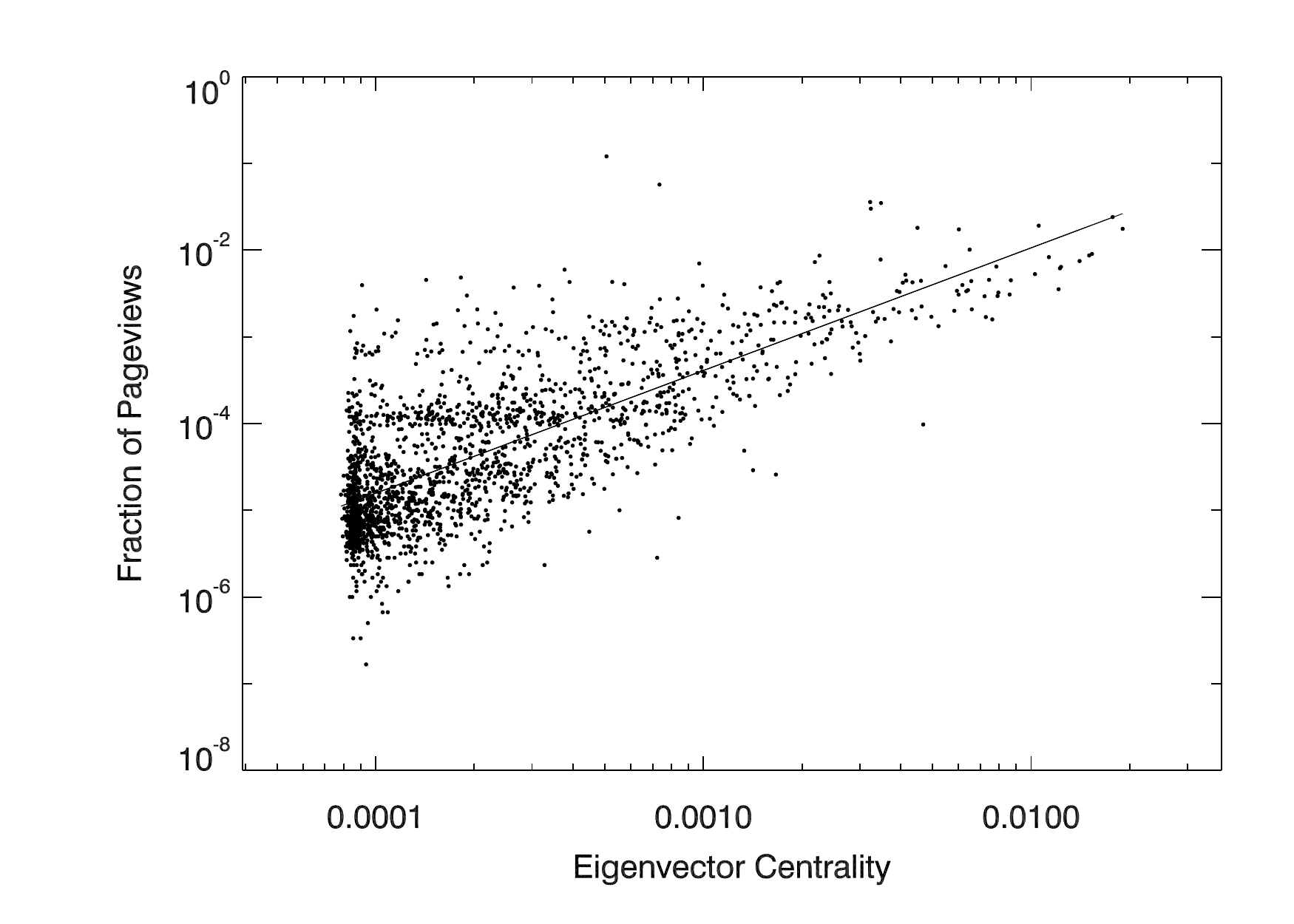}
\caption{The relationship between EC of a page and the percent of the network's page views it accumulates.\label{ec_page_views}}
\end{figure}

\section{Regression on Age and Edits}
\label{regression_section}

To see how a page's intrinsic properties affect its eigenvector centrality (EC) in the final network, we performed a multivariate regression with page age, number of page edits, number of talk page edits, number of editors and page size (in bytes) as predictors of EC. Including pageviews as a predictor does not significantly improve $R^{2}$; we leave it out of our regression model.

We normalized all predictors by z-score to allow comparison between coefficients. We considered two relationships between EC and predictor variables: a linear model and a logistic model. We found the linear model has lower mean-squared error and report the coefficients in Table~\ref{regression_table}.
\begin{table}[H]
\small
\centering
\begin{tabular}{lcc}
\toprule
\textbf{Predictor} & \textbf{Coefficient} \boldmath{$\times10^{-5}$} & $\textbf{\emph{p}}$ \textbf{Value} \\ \hline
Number of editors & $95\pm6$ & $<$$10^{-3}$ \\
Number of talk edits & $46\pm3$ & $<$$10^{-3}$ \\
Page size & $2\pm2$ & n.d. \\
Age & $2\pm2$ & n.d. \\
Number of edits & $-30\pm7$ & $<$$10^{-3}$ \\
\bottomrule
\end{tabular}
\caption{Coefficients of a multivariate linear regression for page eigenvector centrality (EC). The $R^2$ for the fit, including all predictors, is $0.57$. \label{regression_table}}
\end{table}

\begin{figure}[H]
\begin{tabular}{cc}
 \includegraphics[width=2.8in]{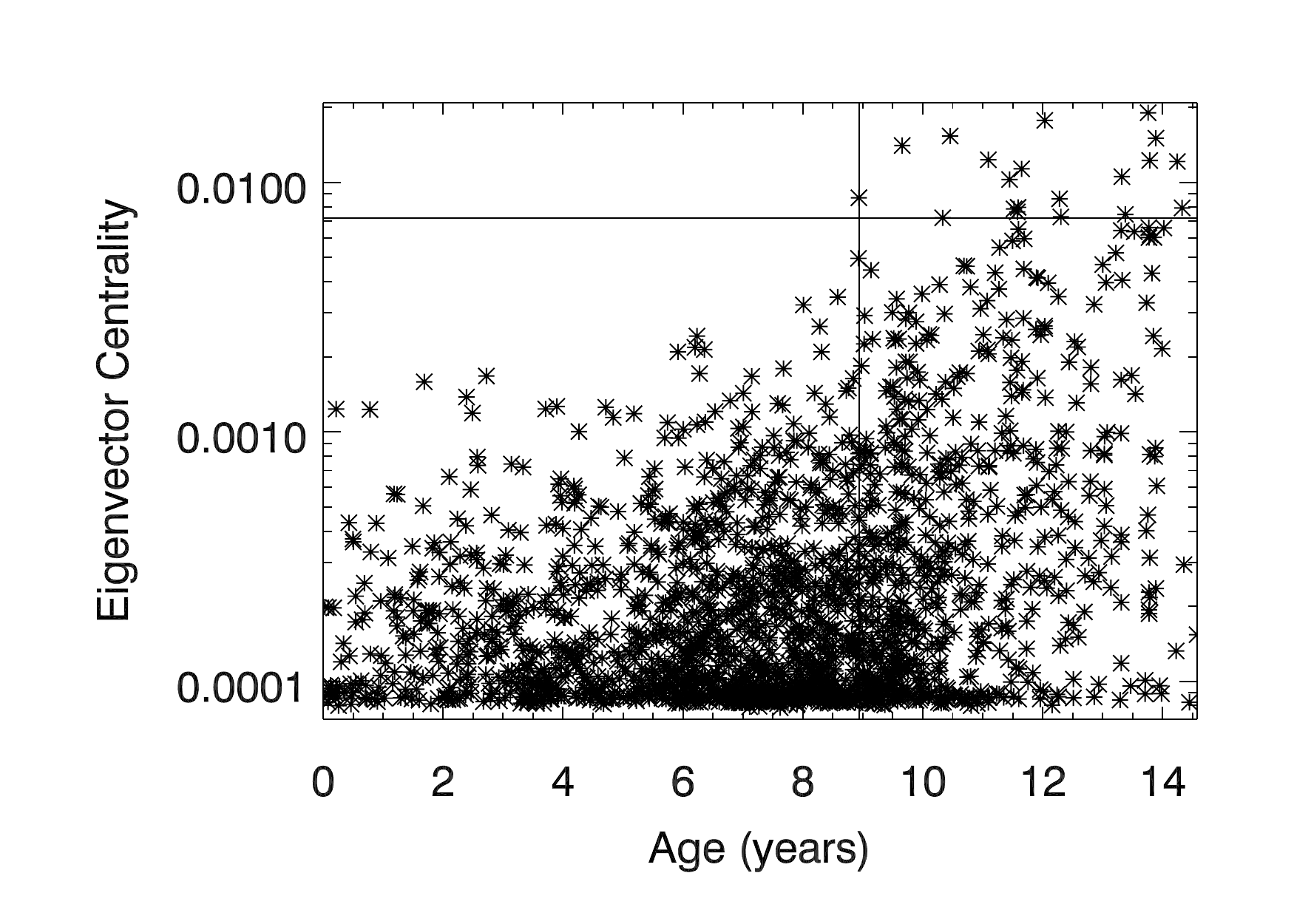} & \includegraphics[width=2.8in]{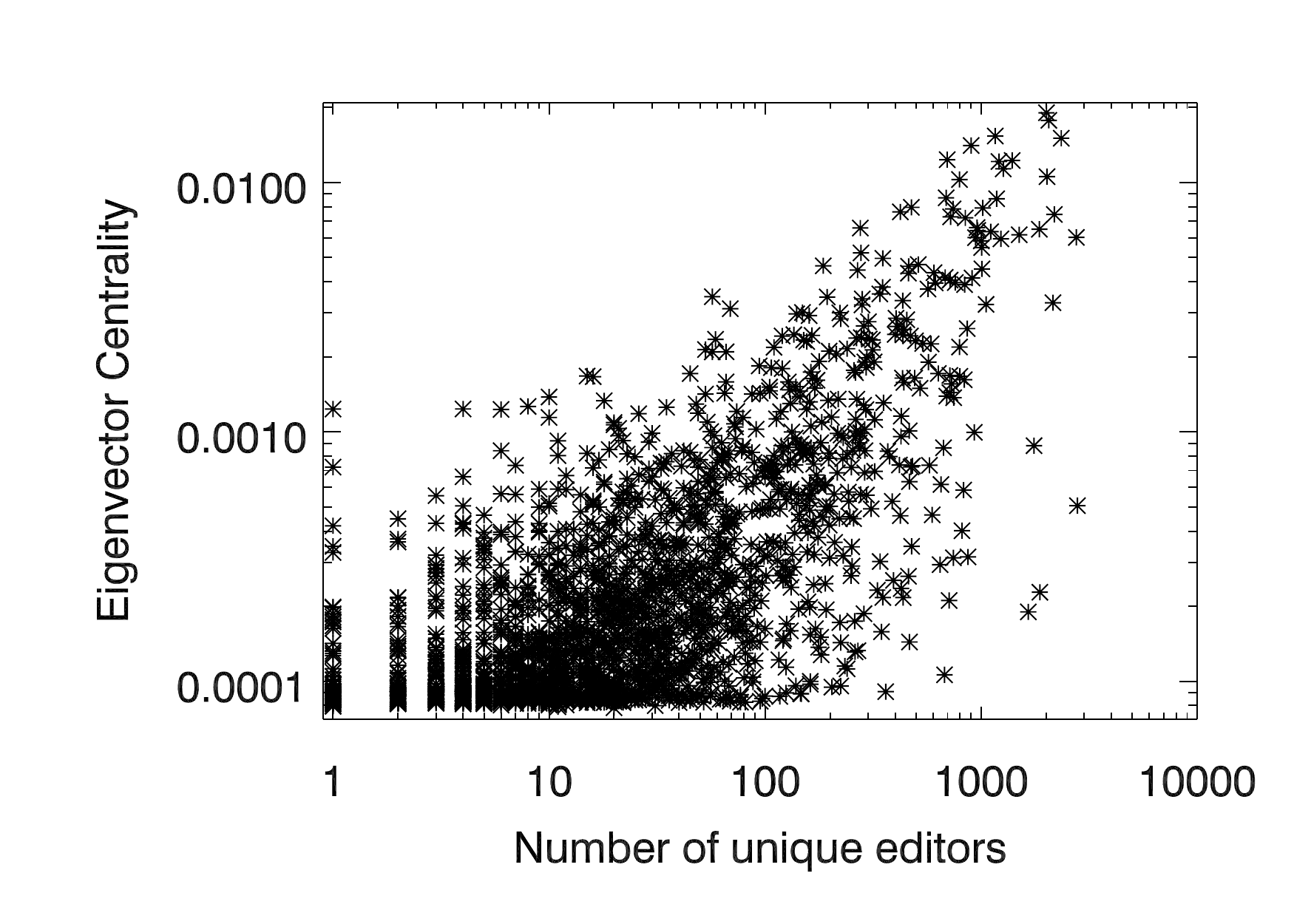} \\
 \end{tabular}
\caption{Important pages are old, but not all old pages are important. {Left} panel: page age (from the end of our data, in August 2015) \emph{vs.} eigenvector centrality; ``core norms'' ({top} twenty pages by EC) are marked by a lower bound in EC and a lower bound in age. While the very top pages in the hierarchy are all old (in the {top-right} region), there are many old pages that have eigenvector centrality comparable to much younger pages. {Right} panel: number of (unique) editors on the page \emph{vs.} eigenvector centrality. A much tighter correlation shows that pages that attract many unique editors have higher EC. When both effects are taken into account in a simple linear regression model, the number of editors dominates.\label{regression_explain}}
\end{figure}

As noted in the main text, our results show that age is a weak predictor of EC, once other variables are included. The number of unique editors is a very strong predictor, as is the number of edits to the talk page. Figure~\ref{regression_explain} 
shows the distinct effect of page age and number of editors. While the most important pages are also the oldest, there are many old pages that are not important at all; the skewed distribution of eigenvector centrality means that this signal is largely washed out in a simple linear model that does not take into account the increasing variance. To reach the top 1\% in EC, you must be old; but to be old is not enough.

By contrast, pages with many editors tend to be high-EC, and there are very few pages with many editors that are not also high in EC. High-EC pages not only attract more page views (see Section~\ref{page_views}), but also more editors. Interestingly, the total number of edits has a negative coefficient in our regression; while there is a strong positive correlation between the number of editors and the number of edits, there are a number of low-EC pages with many edits by a small number of people (e.g., an essay, written in many stages, by a single author, that never gains traction).

\section{Combined Scree Plot}
\label{scree_app}

\begin{figure}[H]
\centering
\includegraphics[width=4.5in]{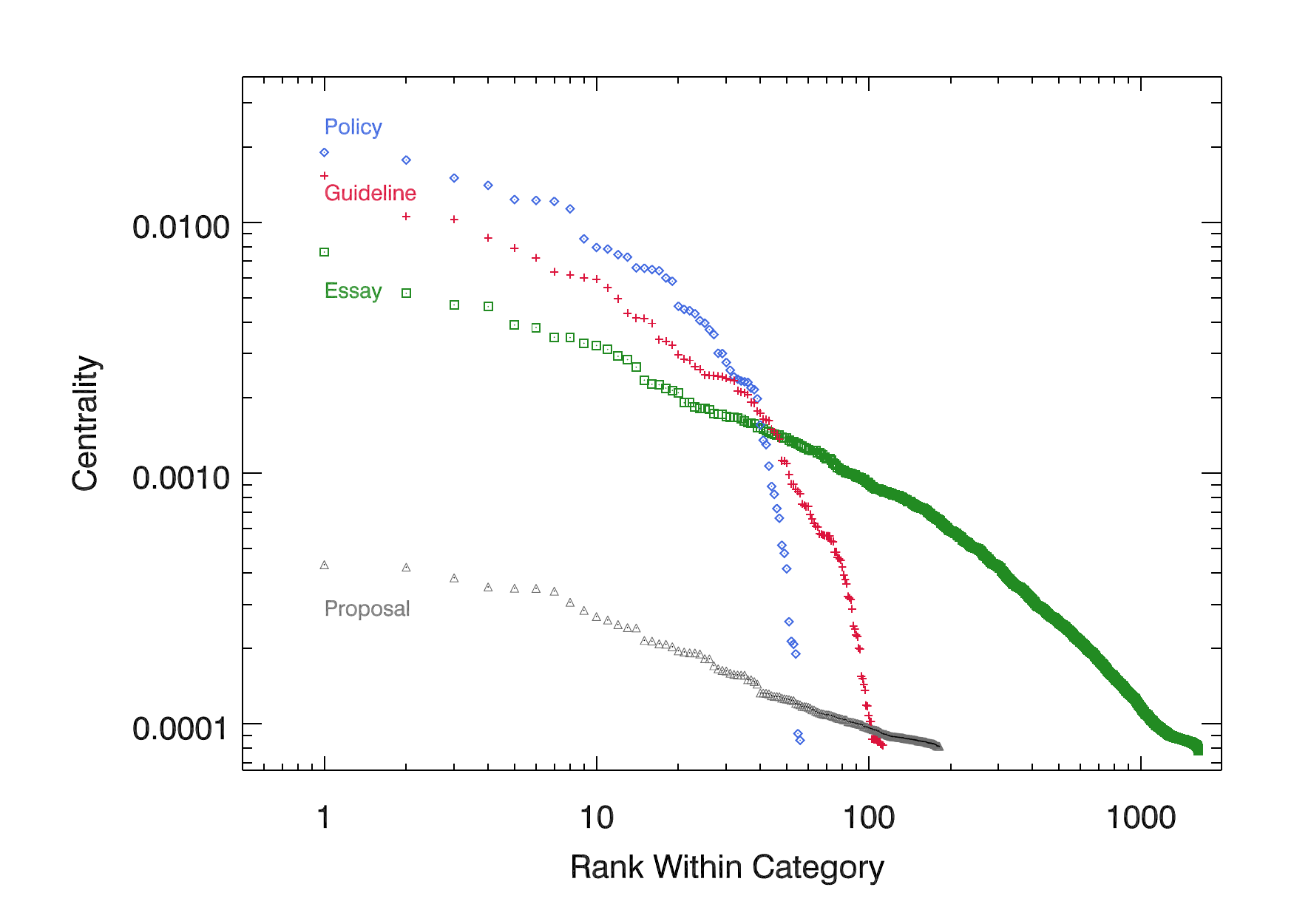}
\caption{Ranked eigenvector centrality for pages, broken out by page category. Policy ({blue} diamond) and guideline ({red} plus) pages dominate the system. More interpretive essays ({green} squares; includes humor and related pages), the most common by number, appear at lower relative rank; the highest ranked essay, for example, has lower centrality than the 10th ranked policy. \mbox{Proposals, failed} or current ({grey} triangles), are the lowest ranked of all.\label{ec}}
\end{figure}

\begin{figure}[H]
\centering
\includegraphics[width=3.5in]{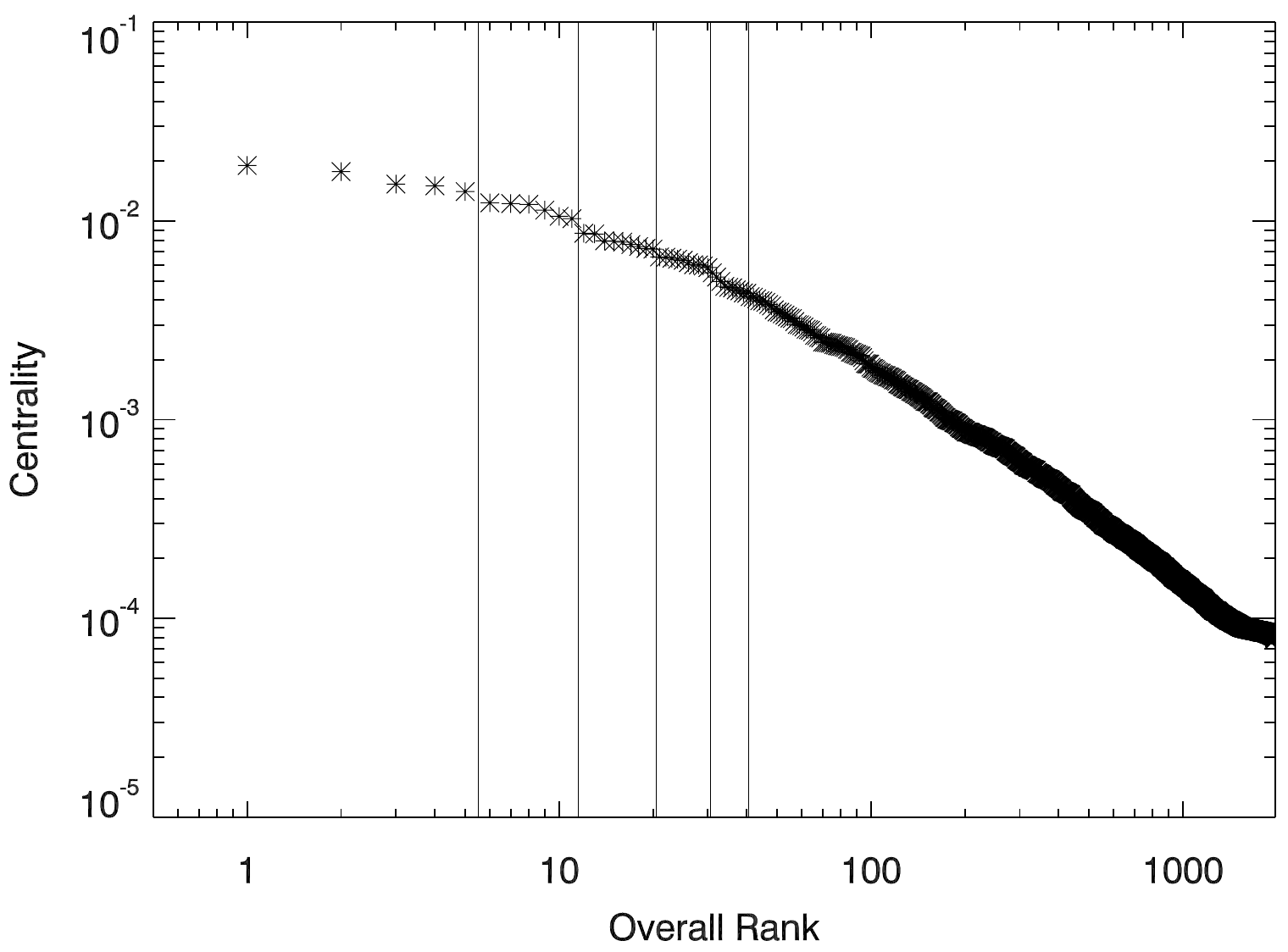}
\caption{Eigenvector centrality for all the pages in our data, ordered by rank. Major divisions (see text) are marked by vertical lines.\label{ec_app}}
\end{figure}

Figure~\ref{ec} shows the rank distribution of EC, page by page, broken out by page class.\mbox{ Defining $E_i$} as the eigenvector centrality of the $i$-th ranked norm allows us to define the break size, $E_{i}-E_{i+1}$, between this norm and the next. Ranking break-sizes allow us to note positions where the remainder of the norms in the system have significantly lower EC. Constraining breaks to be greater than five pages apart leads to the top five divisions shown in Figure~\ref{ec_app}. In the main paper, we list nodes up to the third break-point.

%
%

\section{Local Clustering Coefficient}
\label{app_clustering}

\begin{figure}[H]
\centering
\includegraphics[width=3.5in]{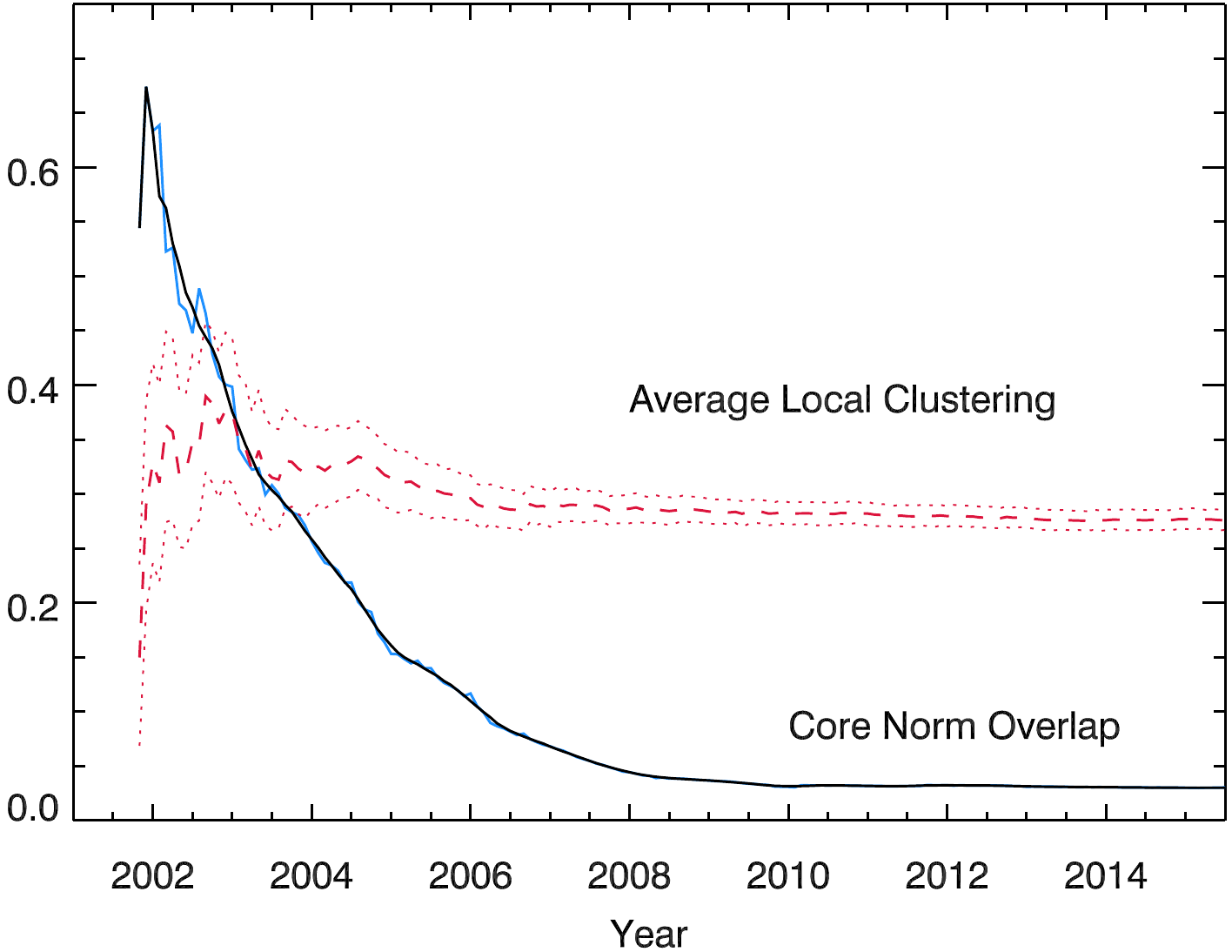}
\caption{The average local clustering coefficient, as a function of time. Despite large-scale changes in overall network properties, this local property remains remarkably constant.}
\label{clustering}
\end{figure}

Our work here focuses on the evolution of global network properties, such as eigenvector centrality, overlap and semantic coherence, that cannot be known by breaking the graph into subgraphs. It is interesting to consider more local measures, however, since these are likely to be under far greater direct user control. The example we consider here is average local clustering, defined as:
\begin{equation}
\kappa=\sum_{i\in G} \frac{\sum_{j,k\in\mathcal{N}(i)} \delta_{jk}}{|\mathcal{N}(i)||\mathcal{N}(i)-1|}
\end{equation}
or, in other words, the number of edges connecting nodes in the neighborhood of $i$, as a fraction of the total number of possible connections between those neighbors. If individuals have a tendency to connect up the network when they create a new node, by linking together nodes it links to, this will tend to increase the clustering. Figure~\ref{clustering} shows this over time. Despite large changes in both population and network size, clustering remains surprisingly constant, at around one-third.

\section{Clusters and Topic Modeling}
\label{k20}

For our base model with $k=20$ topics, Table~\ref{topic_list} shows the top twenty representative words for each topic; in this table we drop the word ``wikipedia'', plurals (except the word ``wikipedias'') and date/time terms (``january'', ``utc'', \emph{etc.}). We use Jason Adams' software package ``lda-ruby'' 
 package ({\url{https://github.com/ealdent/lda-ruby}}), a ruby wrapper for the C code of David M. Blei; this code estimates model parameters using a variational Expectation Maximization algorithm (\url{http://www.cs.princeton.edu/\~blei/lda-c/}~\cite{Blei2003}). In Table~\ref{topic_list}, we show the topics, and their associated words, ordered by the topic's (word-level) prevalence within the encyclopedia. 

For each page, we can compute a distribution over topics; this is just the average of the word-level distributions. By averaging these topic distributions over pages, we can compute the topic distribution for each Louvain community (Collaboration, Article Quality, \textit{etc}.). It turns out that each of the top eight communities has a different most-common topic. This allows us to associate some of the topics we find with a particular cluster, and we list this correspondence in column three of Table~\ref{topic_list}. \mbox{Inspection of} the representative words for these eight topics provides complementary evidence in favor of the community labels, which were previously chosen by manual inspection of the top ten pages by eigenvector centrality (Table~\ref{topic_ec}). 

\newpage
\paperwidth=\pdfpageheight
\paperheight=\pdfpagewidth
\pdfpageheight=\paperheight
\pdfpagewidth=\paperwidth
\newgeometry{layoutwidth=297mm,layoutheight=210 mm, left=2.7cm,right=2.7cm,top=1.8cm,bottom=1.5cm, includehead,includefoot}
\fancyheadoffset[LO,RE]{0cm}
\fancyheadoffset[RO,LE]{0cm}

\begin{table}[H]
\centering
\small
\scalebox{0.8}[0.8]{
\begin{tabular}{ccll}
\toprule
\textbf{Rank} & \textbf{Fraction} & \textbf{Louvain Community} & \textbf{Representative Words} \\ \midrule
\multirow{2}{*}{1} & \multirow{2}{*}{11.4\%} & \multirow{2}{*}{Collaboration} & editor, edit, dont
, good, people, make, editing, policy, page, talk, \\
  & & & time, article, faith, point, policies, encyclopedia, consensus, community, personal, user \\ \hline
\multirow{2}{*}{2} & \multirow{2}{*}{8.67\%} & \multirow{2}{*}{Article Quality} & source, reliable, article, material, information, research, primary, view, original, editors, \\
  & & & subject, published, secondary, policy, neutral, point, scientific, content, topic, claims \\ \hline
\multirow{2}{*}{3} & \multirow{2}{*}{8.56\%} & \multirow{2}{*}{---} & article, deletion, page, deleted, discussion, content, delete, speedy, talk, tag, \\
  & & & subject, information, policy, user, guidelines, criteria, notability, afd
	, time, essay\\\hline
\multirow{2}{*}{4} & \multirow{2}{*}{8.26\%} & \multirow{2}{*}{Experts and Credentials} & article, information, content, encyclopedia, editors, people, wikipedias, subject, featured, quality, \\
  & & & good, list, topic, readers, time, work, project, knowledge, number, lead \\ \hline
\multirow{2}{*}{5} & \multirow{2}{*}{6.65\%} & \multirow{2}{*}{---} & consensus, policy, discussion, community, process, committee, arbitration, editors, administrator, user, \\
  & & & request, policies, admin, block, dispute, page, wikimedia, proposal, information, made\\\hline
\multirow{2}{*}{6} & \multirow{2}{*}{5.80\%} & \multirow{2}{*}{Formatting Articles} & article, names, title, page, english, disambiguation, naming, redirect, conventions, common, \\
  & & & term, style, citation, word, language, topic, book, usage, examples, cases \\ \hline
\multirow{2}{*}{7} & \multirow{2}{*}{5.69\%} & \multirow{2}{*}{Administrators} & user, edit, page, vandalism, account, ip, editing, talk, editors, bot, \\
  & & & address, protection, administrators, userboxes, username, blocked, block, request, sock, template \\ \hline
\multirow{2}{*}{8} & \multirow{2}{*}{5.36\%} & \multirow{2}{*}{---} & notable, article, notability, list, sources, coverage, criteria, information, subject, reliable, \\
  & & & emojif
	, guideline, film, event, university, significant, general, topic, independent, inclusion\\\hline
\multirow{2}{*}{9} & \multirow{2}{*}{5.03\%} & \multirow{2}{*}{---} & page, link, text, image, file, wikimedia, search, commons, web, information, \\
  & & & external, software, content, article, site, add, click, wiki, edit, make\\\hline
\multirow{2}{*}{10} & \multirow{2}{*}{4.38\%} & \multirow{2}{*}{---} & talk, edit, page, user, war, im, article, dont, people, time, \\
  & & & contribs, good, contributions, back, long, list, things, make, day, ive
	\\\hline
\multirow{2}{*}{11} & \multirow{2}{*}{4.04\%} & \multirow{2}{*}{Content Policies} & copyright, image, public, nonfree, free, work, content, license, domain, law, \\
  & & & fair, article, copyrighted, published, states, pma, united, subject, permission, media \\ \hline
\multirow{2}{*}{12} & \multirow{2}{*}{3.81\%} & \multirow{2}{*}{---} & page, talk, template, namespace, user, link, article, text, category, section, \\
  & & & special, edit, title, list, signature, ut
	, mediawiki, redirect, move, navbox\\\hline
\multirow{2}{*}{13} & \multirow{2}{*}{3.28\%} & \multirow{2}{*}{Humor} & list, chart, people, united, war, town, world, man, england, states, \\
  & & & british, top, hot, songs, women, city, ireland, music, number, death \\ \hline
\multirow{2}{*}{14} & \multirow{2}{*}{3.04\%} & \multirow{2}{*}{---} & category, day, categories, article, tip, stub, list, page, people, categorization, \\
  & & & main, link, year, created, red, featured, create, template, sort, subcategories\\\hline
\multirow{2}{*}{15} & \multirow{2}{*}{2.97\%} & \multirow{2}{*}{Wiki-larping} & people, user, time, status, wikidragon, truth, wikifauna, wikipuma, credentials, names, \\
  & & & editathon, work, turkish, years, page, make, real, history, group, greek \\ \hline
\multirow{2}{*}{16} & \multirow{2}{*}{2.83\%} & \multirow{2}{*}{---} & support, oppose, policy, people, user, proposal, talk, userboxes, dont, image, \\
  & & & offensive, pov
	, namespace, page, content, article, censorship, vote, npov
	, agree\\\hline
\multirow{2}{*}{17} & \multirow{2}{*}{2.79\%} & \multirow{2}{*}{---} & ban, topic, editing, indefinite, talk, sanctions, article, page, user, edit, \\
  & & & discussion, banned, paid, related, editor, contribs
	, interest, coi
	, community, broadly\\\hline
\multirow{2}{*}{18} & \multirow{2}{*}{2.55\%} & \multirow{2}{*}{---} & quotation, style, citing, punctuation, american, mos, ads, dash, manual, inactive, \\
  & & & en
	, english, issue, sentence, dashes, election, text, space, british, jumped\\\hline
\multirow{2}{*}{19} & \multirow{2}{*}{2.41\%} & \multirow{2}{*}{Page Templates} & text, template, page, line, article, gt, section, lt, enforcement, table, \\
  & & & footnote, law, summary, infobox, style, agencies, synth
	, color, work, data \\ \hline
\multirow{2}{*}{20} & \multirow{2}{*}{2.32\%} & \multirow{2}{*}{---} & article, station, number, year, state, route, highway, time, road, points, \\
  & & & date, railway, britannica, ship, include, information, eb
	, county, class, official\\\bottomrule
\end{tabular}}

\caption{Representative one-grams from each of the topics in our $k=20$ topic model, ranked by the weighted fraction of words assigned. The top nine Louvain clusters are each dominated by a unique topic. \label{topic_list}}

\end{table}

\begin{table}[H]
\centering
\small
\scalebox{0.8}[0.8]{
\begin{tabular}{cll}
\toprule
\textbf{Rank} & \textbf{Cluster Name} & \textbf{Top Pages} \\ \midrule
1 & Article Quality & Neutral\_point\_of\_view; Verifiability; Identifying\_reliable\_sources; What\_Wikipedia\_is\_not; Biographies\_of\_living\_persons; No\_original\_research; Citing\_sources \\
2 & Collaboration & Consensus; Policies\_and\_guidelines; Assume\_good\_faith; Dispute\_resolution; Civility; Edit\_warring; Talk\_page\_guidelines \\
3 & Administrators & Administrators; Blocking\_policy; Arbitration\_Committee; Vandalism; User\_pages; Sock\_puppetry; User\_access\_levels \\
4 & Formatting Articles & Redirect; Article\_titles; Disambiguation; Manual\_of\_Style; Namespace; What\_is\_an\_article?; Categorization \\
5 & Content Policies & Copyrights; Copyright\_violations; Non-free\_content; Image\_use\_policy; General\_disclaimer; Non-Wikipedia\_disclaimers; Substitution \\
6 & Wiki-larping & Citation\_needed; Wikibreak; WikiGnome; Wikipediholic; Talk\_page\_stalker; Wikipedia\_is\_a\_volunteer\_service; WikiDragon \\
7 & Page Templates & Overlink\_crisis; Pruning\_article\_revisions; Disinfoboxes; Thinking\_outside\_the\_infobox; Advanced\_template\_coding; Advanced\_article\_editing; Advanced\_footnote\_formatting \\
8 & Experts and Credentials & Expert\_editors; Honesty; Expert\_retention; Randy\_in\_Boise; Ten\_Simple\_Rules\_for\_Editing\_Wikipedia; Conflicts\_of\_interest\_(medicine); There\_is\_no\_credential\_policy \\
9 & Humor & Silly\_Things; Rules\_for\_Fools; April\_Fools; April\_Fool's\_Main\_Page; Unusual\_articles; Yet\_more\_Best\_of\_BJAODN; Best\_of\_BJAODN \\
\bottomrule
\end{tabular}}

\caption{Top pages within each cluster, by eigenvector centrality. \label{topic_ec}}
\end{table}

%
\newpage
\restoregeometry
\paperwidth=\pdfpageheight
\paperheight=\pdfpagewidth
\pdfpageheight=\paperheight
\pdfpagewidth=\paperwidth
\headwidth=\textwidth

\bibliographystyle{mdpi}
\renewcommand\bibname{References}

\end{document}